\documentclass[prd,showpacs,nofootinbib]{revtex4}
\usepackage{graphicx}
\usepackage{latexsym}
\usepackage{amsmath}
\usepackage[all]{xy}

\newcommand{\bea}{\begin{eqnarray}}
\newcommand{\eea}{\end{eqnarray}}
\newcommand{\be}{\begin{equation}}
\newcommand{\ee}{\end{equation}}

\begin{document}

\title{Newtonian limit of nonlocal cosmology}
\author{Tomi S. Koivisto}
\email{T.Koivisto@thphys.uni-heidelberg.de}
\affiliation{Institute for Theoretical Physics, University of Heidelberg, 69120 Heidelberg,Germany}

\begin{abstract}

We study the consequences of the $f(R/\Box)$ gravity models for the Solar system and the large scale structure of the universe. 
The spherically symmetric solutions can be used to obtain bounds on the constant and the linear parts of the correction terms.
The evolution of cosmological matter structures is shown to be governed by an effectively time dependent Newton's constant. 
We also analyze the propagation of the perturbation modes. Tensor and vector modes are only slightly modified, 
but two new scalar degrees of freedom are present. Their causality and stability is demonstrated, and their formal ghost conditions are related 
to a singularity of the cosmological background. In general, the Newtonian limit of these models has no apparent conflicts with observations 
but can provide useful constraints.

\end{abstract}  

 
\maketitle

\section{Introduction}

One covariant way of generalizing gravity is to consider the effective Newton's constant as a function of $\phi=R/\Box$, the inverse 
d'Alembertian acting on the scalar curvature. This kind of nonlinear dependence can be parameterized as the following 
\be \label{action}
S = \frac{1}{2\kappa^2}\int d^4 x \left(1+f(\frac{R}{\Box})\right)R.
\ee
General relativity is the limit $f=0$. 
This form of the action can be motivated\footnote{On the other hand, such models might appear in (stringy) compactification scenarios. This will 
become clearer when we introduce an equivalent description in terms of kinetically coupled scalar fields (non linear sigma model).} by the fact that 
quantum effects generically introduce corrections involving different operators and curvature invariants \cite{Donoghue:1994dn,Shapiro:2008sf}. 
The exact form of the corrections is of course beyond control, but Eq.(\ref{action}) provides a simple
parameterization, which can be constrained by observations: this is the phenomenological approach we take in the present study. 

Previously it has been noted that the most appealing (linear) form, $f(\phi) \sim \phi$, could cure the unboundedness of the Euclidean gravity action 
\cite{Wetterich:1997bz}. More recently it was suggested \cite{Deser:2007jk} that these types of effective gravity actions could incorporate the observed 
cosmic acceleration with reduced fine tuning. The problem of the magnitude of the cosmological constant might be evaded since $\phi = 
R/\Box$ is a dimensionless combination, and thus no new scales (in addition to the Planck scale) need to be introduced. The coincidence problem could 
also get a simple resolution, since $R$ and thus $\phi$ vanish during radiation domination, so the acceleratory effect can be interpreted as a 
delayed consequence of the onset of the matter dominated era.
  
Detailed analysis of the background dynamics shows that this could work, even for the simplest forms of $f$ \cite{Koivisto:2008xf}. Unresolved issues 
exist though, like an occurrence of a sudden future singularity \cite{Barrow:2004xh,Bamba:2008ut}. In addition, in the simplest models featuring the late 
acceleration, one finds large effects taking place also during inflation. On the other hand, this could be used as beneficial effect to end inflation 
and reheat to universe, or even to drive inflation without the usual scalar fields as attempted in Ref.\cite{Nojiri:2007uq}. Realizing this possibility would 
require to find the classes of actions unifying consistently very early and the late universe. The fact that the action (\ref{action}) is capable of 
reproducing an arbitrary expansion history \cite{Koivisto:2008xf}, hints that this is possible. However, in this work omit the background problematics, since 
our aim is to focus on inhomogeneous evolution of these models, particularly at the important Newtonian limit.   

This limit within the $f(R)$ class of models has been intensively studied in connection with the dark matter and dark energy problems, 
see \cite{Nojiri:2006ri,Sotiriou:2008rp,Sotiriou:2007yd,Capozziello:2007ec,Lobo:2008sg} for reviews. Even if viable examples would be found in that 
class of models, there seems to be no reason to prefer those more complicated models to the simple cosmogical constant, since they have not (at least 
this far) promised any help with the cosmological fine tuning problems. In this regard the $f(R/\Box)$ scenarios make progress, as they, like 
mentioned above, could provide both a mechanism to connect the onset of the acceleration with the past matter dominated era, and remove the need 
for unnatural constants in the theory. Though also a $f(R)$ action can be recasted into a scalar-tensor theory, the structure of the action 
(\ref{action}), which we'll soon clarify, is different. 

There has also been recently much interest in nonlocal scalar field cosmologies \cite{Barnaby:2007ve}. These are motivated by the p-adic tachyon and
open string field theory actions, which include operators with infinite number of derivatives acting on scalar fields. In particular, the exponent of 
the d'Alembertian, $e^\Box$, appears. These kind of theories are equivalent to theories with several (possibly an infinite number of) coupled scalar fields, 
as soon becomes explicit in a different context. Nonperturbative methods can then be applied to treat localizable systems via higher-dimensional
formulation \cite{Calcagni:2007ru,Calcagni:2007ef,Calcagni:2008nm,Mulryne:2008iq}.  
Friedmann-Robertson-Walker (FRW) solutions have been studied in string field theory 
\cite{Aref'eva:2004vw,Aref'eva:2007uk,Calcagni:2005xc,Barnaby:2008fk}, and in related string-inspired forms of nonperturbative gravity with 
infinitely many derivatives \cite{Biswas:2005qr,Biswas:2006bs}. 

Finally, let us remark that in the so called degravitation models the effective Newton's constant is promoted to a nonlocal operator, whose filtering 
effect 
could explain why gravity is insensitive to a cosmological constant \cite{ArkaniHamed:2002fu}, leaving the possibility of the present cosmological term 
appearing as an afterglow of inflation \cite{Patil:2008sp}. These interesting models are burdened with problems associated to massive 
gravity \cite{Dvali:2007kt}, including the 
loss of covariance, as the change of $G$ is done at the level of field equations and not of the action. Though they are thus fundamentally different types of 
gravity modifications than we consider in the present paper, one can observe some analogous workings at a phenomenological level. In fact, similar
actions could be constructed approximating the degravitation models \cite{Barvinsky:2003kg}. 

The present paper is devoted to study of inhomogeneous solutions of the $f(\phi)$ case, and in particular in the weak field limit where the 
inhomogeneities appear as perturbations about more symmetric solutions. We study all types of cosmological perturbations for the first time, and look 
at the Solar system limit in more detail than previously done. In section (\ref{sektio}) we present the model in generalized framework. In 
section (\ref{sektio2}) we write down the equations for each type of linear perturbations. The post-Newtonian limit, relevant for the Solar system 
physics and for the evolution of cosmological structure, is determined in section (\ref{sektio3}). The propagation of the scalar modes is more 
complicated than that of the vector and tensor modes, and we discuss that 
in the section (\ref{sektio4}). The conclusion is made in section (\ref{sektio5}).

\section{The model and its generalizations}
\label{sektio}

This section prepares for the study of perturbations.
We present an equivalent description of action (\ref{action}) as a biscalar-tensor theory, that will be employed throughout the paper. We do this in a 
wider framework, looking first at more general operators than the d'Alembertian and then at more general functions $f$ which may depend on several
variables. Finally we write down the cosmological background system.  

\subsection{Biscalar-tensor representation}

We consider the following class of nonlocal gravity actions,
\be
S  =  \frac{1}{2\kappa^2}\int d^4 x \sqrt{-g}R\left(1+f(\Delta^{-1}R)\right). \label{eka}  
\ee
where $\Delta$ is some operator of the form $\Delta = \nabla_\mu Q^{\mu\nu} \nabla_\nu$. 
A simple example is the d'Alembertian, $\Box = \nabla_\mu\nabla^\mu$. A more complicated 
example is 
\be
\Delta_P = \Box + \nabla_\mu (\alpha R^{\mu\nu} - \beta Rg^{\mu\nu})\nabla_\nu,
\ee
When $\alpha=1$ and $\beta=1/3$, one has the Paneitz operator which is known to arise in conformal anomalies \cite{Deser:1999zv}. 
Einstein tensor -type model corresponds to $\beta=\alpha/2$, and the box model is recovered when $\alpha=\beta=0$.   
%\cite{Deser:2007jk,Nojiri:2007uq}.
Introducing the Lagrange multiplier $\xi$ to rename the inverse of $\Delta$ acting on R as $\phi$, we rewrite the action in the form 
\be
S =  \frac{1}{2\kappa^2}\int d^4 x \sqrt{-g}\left[(1+f(\phi))R + \xi(R-\Delta\phi)\right].
\ee
Thus one may consider the theory as a multiscalar-tensor theory \cite{Damour:1992we}. It is convenient to introduce a new field, $\psi = f(\phi)-\xi$, 
which is then the only one with a non-minimal gravity coupling. After a partial integration, the action assumes the form
\be \label{aktio1}
S =  \frac{1}{2\kappa^2}\int d^4 x \sqrt{-g}\left[(1+\psi)R + Q^{\mu\nu}\left(\nabla_\mu \psi - f'(\phi)\nabla_\mu\phi\right)\nabla_\nu\phi\right].
\ee
When this is coupled to the matter action, we get the field equations
\bea
G_{\mu\nu} = \frac{1}{2\kappa^2}T^m_{\mu\nu} + T^\phi_{\mu\nu} + T^\psi_{\mu\nu} + T^X_{\mu\nu},  \label{efe}
\eea
where $T^m_{\mu\nu}$ is the matter energy momentum tensor. The effective energy momentum tensors for the two fields and for their coupling read as
\bea
T^\phi_{\mu\nu} & = & 
\left[-\frac{1}{2}Q^{\alpha\beta}g_{\mu\nu} + \Sigma_{\mu\nu}^{\phantom{\mu\nu}\alpha\beta}\right]f'(\phi)
(\nabla_\alpha\phi)(\nabla_\beta\phi),  \label{set_phi} \\
T^\psi_{\mu\nu} & = & -\left(G_{\mu\nu} + g_{\mu\nu}\Box-\nabla_\mu\nabla_\nu \right)\psi,                          \label{set_psi} \\ 
T^X_{\mu\nu} & = & \left[\frac{1}{2}Q^{\alpha\beta}g_{\mu\nu} - \Sigma_{(\mu\nu)}^{\phantom{\mu\nu}\alpha\beta}\right]
(\nabla_\alpha\phi)(\nabla_\beta\psi).  \label{set_x}
\eea 
Here $\Sigma_{\mu\nu}^{\phantom{\mu\nu}\alpha\beta}$ is a derivative operator, which includes derivatives up to $m$'th order, if $Q_{\mu\nu}$ involves
$m$'th order derivatives of the metric. More explicitly we can write
\be
\Sigma_{\mu\nu}^{\phantom{\mu\nu}\alpha\beta} = \sum_{k=0}^m\sum_{n=k}^{m}(-1)^n{n \choose k}\left[\nabla_{\gamma_{k+1}} \ldots \nabla_{\gamma_n}
\left(\frac{\partial Q^{\alpha\beta}}{\partial g^{\mu\nu}_{,\gamma_1 \ldots \gamma_n}}\right)\right]\nabla_{\gamma_1} \ldots \nabla_{\gamma_k}.
\ee  
In the simplest case $Q^{\alpha\beta}=g^{\alpha\beta}$, we have $\Sigma_{\mu\nu}^{\phantom{\mu\nu}\alpha\beta} = \delta_\mu^\alpha\delta_\nu^\beta$.
Only the sum of $T^\phi_{\mu\nu}+T^\psi_{\mu\nu}+T^X_{\mu\nu}$ is separately conserved. This implies the generalized Bianchi identity and
usual conservation of the matter energy momentum tensor \cite{Koivisto:2005yk}, $\nabla^\mu T^m_{\mu\nu}=0$.

\subsection{More general models}

We consider a generalization of the model with an arbitrary number of derivatives. Let us then write 
\be
S  =  \frac{1}{2\kappa^2}\int d^4 x \sqrt{-g}R\left(1+f(\Delta^{-m}R,\dots,\Delta^{-1}R,\Delta R,\ldots, \Delta^n R)\right). \label{eka1}
\ee
This describes a large amount of possible models. The nature of these models is though not very transparent in this form. To better 
understand the field content and couplings of the model, we rewrite it in terms of scalar fields.
Proceeding as previously, we arrive at the action
\be \label{aktio2}
S  =  \frac{1}{2\kappa^2}\int d^4 x \sqrt{-g}\left[(1+\psi)R + Q^{\mu\nu}(\nabla_\mu\chi_1)(\nabla_\nu R) + V + K\right],
\ee
where the potential term is
\be \label{pot}
V = -\sum_{k=2}^m \phi_{k-1}\xi_k + \sum_{k=1}^n \varphi_k\chi_k,
\ee
and the kinetic term has the form
\bea
K = Q^{\mu\nu}\Bigg[(\nabla_\mu\phi_1)(\nabla_\nu\psi)-\frac{\partial f}{\partial \phi_1}(\nabla_\mu\phi_1)(\nabla_\nu\phi_1)
& - & \sum_{k=2}^m (\nabla_\mu\phi_k)\left(\frac{\partial f}{\partial \phi_k}(\nabla_\nu\phi_1) + (\nabla_\nu\xi_k)\right) \nonumber \\
& + & \sum_{k=1}^n (\nabla_\mu\varphi_k)\left(\frac{\partial f}{\partial \varphi_k}(\nabla_\nu\phi_1) + (\nabla_\nu\chi_{k+1})\right)
\Bigg],
\eea
where $f$ is a function of the $n+m$ scalar fields, $f=f(\phi_m,\dots,\phi_1,\varphi_1,\dots,\varphi_n)$. Nonlocality of the theory means that 
either $m$ or $n$ goes to infinity, and in a genuinely nonlocal one cannot reduce the number of derivatives (i.e. recast it in a local form); it is 
however 
common terminology to call e.g. the $f(R/\Box)$ nonlocal. It is now clear that each higher power of  
derivative and inverse derivative in the action (\ref{eka1}) adds two scalar degrees of freedom into the theory. At face value it seems that $n+m+1$ 
of these are non-minimally coupled to gravity, however, we have shown that any nonzero $n$ or any nonzero $m$ both in essence add just one 
non-minimally coupled scalar into the play. In the Einstein frame the coupling resulting from nonzero $m \neq 0$ is transformed into non-minimal 
matter coupling. 

In passing, we remark that changing $\Delta R$ to $\Delta \mathcal{R}$, where $\mathcal{R}$ is some scalar quantity, would change only the 
$\psi R$ terms into $\psi \mathcal{R}$ terms in the actions (\ref{aktio1}) and (\ref{aktio2}). Any scalar $\mathcal{R}$ which vanishes in de Sitter
spaces might be useful, if one wanted to protect the de Sitter type solutions.

It would be interesting to study in which cases the solutions may assume form of converging series. One of the interesting cases is the form 
$f=\log{\Delta}R$, which represents the leading nonlocal loop correction to gravity as an effective quantum field theory (for previous approximative
solutions, see \cite{Espriu:2005qn,Cabrer:2007xm}). In the rest of the present paper, we will however specialize to the $m=1$, $n=0$ case\footnote{The 
generalization 
to the higher-derivative models is not completely straightforward. The potential (\ref{pot}) has to be taken into account whenever $m>1$ or $m>0$. If 
$n$ remains zero, however, one then needs only to generalize all the $\phi$ -dependent kinetic terms.}.

\subsection{Cosmological equations}

Consider the inverse box model $f=f(R/\Box)$. In a flat FRW background,
\be \label{frw}
ds^2 = -a^2(\tau) \left(d\tau^2 + d\mathbf{x}^2\right),
\ee
we get the Friedmann equations 
\be
3H^2(1+\psi) = \frac{a^2 \rho_m}{\kappa^2} + \frac{1}{2}\left(f'(\phi)\dot{\phi}^2 - \dot{\psi}\dot{\phi}\right) - 3H\dot{\psi}
\ee
\be
-(2\dot{H}+H^2)(1+\psi) = \frac{a^2 p_m}{\kappa^2} + \frac{1}{2}\left(f'(\phi)\dot{\phi}^2 - \dot{\psi}\dot{\phi}\right) + \ddot{\psi} + H\dot{\psi}
\ee
Thus the time derivatives of the scalar fields contribute extra energy sources, and the field $\psi$ also modulates effective Newton's constant.  
Note that here dot means a derivative with respect ot the conformal time $\tau$, and $H \equiv (d a/d\tau)/a$ is the conformal Hubble parameter.
The equations of motion for the fields are then
\be
\ddot{\phi}+2H\dot{\phi}=-6(\dot{H}+H^2),
\ee
\be
\ddot{\psi}+2H\dot{\psi} = f''(\phi)\dot{\phi}^2 -12f'(\phi)(\dot{H}+H^2).
\ee
It will be useful to define the energy density fraction of matter as usual,
\be 
\Omega_m \equiv \frac{a^2\rho_m}{3\kappa^2 H^2}.
\ee
The background expansion of these models has been considered elsewhere \cite{Jhingan:2008ym, Nojiri:2007uq, Koivisto:2008xf}. 
In particular, a power-law model $f=f_n \phi^n$ can feature modifications of the expansion both at early 
and at late late times, depending on the sign of $(-1)^n f_n$. Accelerating and super-accelerating scenarios 
exist both in vacuum and in presence of matter. An exponential model $f=f_e e^{\lambda\phi}$ has scaling 
solutions, thus the modification could be present at all times \cite{Koivisto:2008xf}. In the following 
we will not specify any particular evolution but consider the general weak-field limit phenomenology of these models. 
 
%We then have a complete set of dimensionless variables. The dynamical system may then be written as  
%\bea \label{om_evol}
%\Omega_m^* & = & 3(w_m-w_{eff})\Omega_m, \\ \label{psi_evol}
%\Psi^{**} & = & \frac{3}{2}(w_{eff}-1)\Psi^* - 6f'(\phi)(1-3w_{eff}) + 4f''(\phi)X^2, \\ \label{x_evol}
%X^* & = & \frac{3}{2}\left[(w_{eff}-1)X  + 3w_{eff}-1\right].
%\eea
%where
%\be
%w_{eff} = \frac{3w_m\Omega_m  + 2f'(\phi)(X^2-3) + 4f''(\phi)X^2 - (1+X)\Psi^*}{3(1+\Psi-6f'(\phi))}.
%\ee
%The first equation simply expresses the usual matter conservation law, and the two following equations are the conservation laws for
%our two scalar fields. 

\section{Perturbations}
\label{sektio2}

In this section we show the equations governing the evolution of cosmological perturbations. The vector and tensor perturbations are simple, but the 
analysis
of the scalar perturbations, sourced both by the scalar fields and by matter, is continued in later sections. For cosmological perturbations in wide
range of other modified gravity models, see \cite{Hwang:2005hb}. 

Introducing small perturbations about the background, one may write the metric as
\be \label{newt}
ds^2 = a(\tau)^2\left[-d\tau^2(1+2\Psi) + d\tau d{\bf x}\cdot{\bf W}  +  dx_i dx^i(1-2\Phi + H_{ij})\right].
\ee
This is completely general parameterization of the inhomogeneities and anisotropies. Those can be decomposed into scalar, vector 
and tensor parts according the transformation properties under spatial rotations \cite{Mukhanov:1990me}. At linear order, the different 
types of modes decouple. With the scalar perturbations, it is conventional to work in the Newtonian gauge \cite{Bardeen:1980kt}. 
That is specified by the metric potentials
$\Phi$ and $\Psi$. Note that we use the uppercase symbols for the metric potentials (and for all metric perturbations for consistency), 
which have nothing to do with the $\phi$ and $\psi$ which label our scalar fields. The vector type perturbations are characterized by the ${\bf W}$, which is transverse 
and thus has two independent components. The transverse and traceless $H_{ij}$ finally describes the gravitational wave. By construction, it is
gauge invariant and includes two independent polarizations, which are conventionally called $H_+$ and $H_X$. In total, we then have 
6 degrees of freedom in the perturbations, matching with the $D(D+1)/2 - D$ physical degrees of freedom of the metric in dimension 
$D=4$. The metric Eq.(\ref{newt}) is thus completely general and fixes the gauge uniquely as well.

\subsection{Scalar equations}

For completeness we list the equations for scalar perturbations here. 
The ADM energy constraint ($G^0_0$ component of the field equation) is
  \bea \label{admenergy}
  -3\left(H + \frac{\dot{\psi}}{2(1+\psi)} \right)\dot{\Phi} 
   +  \nabla^2\Phi  +  \left[-3H + \frac{1}{2(1+\psi)}\left(f'\dot{\phi^2}-\dot{\phi}\dot{\psi}-6H\dot{\psi}\right)\right]\Psi 
\nonumber \\
  =  \frac{1}{2(1+\psi)}\left[\frac{a^2}{\kappa^2}\delta \rho_m + \left(f'\dot{\phi}-\frac{1}{2}\dot{\psi}\right)\delta\dot{\phi} + 
\frac{1}{2}f''\dot{\phi}^2\delta{\phi} - \left(3H+\frac{1}{2}\dot{\phi}\right)\delta\dot{\psi} + \left(3\dot{H}+\nabla^2\right)\delta\psi\right],
 \eea
and the momentum constraint ($G^0_i$ component) is
 \be \label{mc}
  3\dot{\Phi} + \left(3H + \frac{3}{2}\frac{\dot{\psi}}{1+\psi}\right)\Psi = 
  \frac{3}{2(1+\psi)}\left[\frac{a^2}{\kappa^2}(\rho_m+p_m)\frac{v_m}{k} 
  + \left(f'\dot{\phi}-\frac{1}{2}\dot{\psi}\right)\delta\phi + \delta\dot{\psi} - \left(H+\frac{1}{2}\dot{\phi}\right)\delta\psi\right].
 \ee
The shear propagation equation ($G^i_j-\frac{1}{3}\delta^i_jG^k_k$ component) reads
 \be
 \label{propagation}
  \Phi - \Psi = \frac{1}{1+\psi}\left(\frac{a^2}{\kappa^2}\pi_m+\delta\psi\right),
  \ee
where $\pi_m$ is the matter anisotropic stress.
The Raychaudhuri equation ($G^k_k-G^0_0$ component) is now given by
 \bea\label{ray}
  3\ddot{\Phi}  & + &
 3\left(H + \frac{\dot{\psi}}{2(1+\psi)}\right)\dot{\Phi}
 +3\left(H - \frac{\dot{\psi}}{2(1+\psi)}\right)\dot{\Psi}
   +  \left[6\dot{H} + \frac{1}{2(1+\psi)}\left(6\ddot{\psi} + 2f'\dot{\phi}^2 - 2\dot{\phi}\dot{\psi}\right) + \nabla^2\right]\Psi
 \nonumber \\
 & = &
 \frac{1}{2(1+\psi)}\left[\frac{a^2}{\kappa^2}\left(\delta\rho_m+3\delta p_m\right) + \left(2f'\dot{\phi}-\dot{\psi}\right)\dot{\delta\phi} 
  + f''\dot{\phi}^2\delta\phi + 3\ddot{\psi} - \left(3H+\dot{\phi}\right)\delta\dot{\psi} - \left(6H^2+\nabla^2\right)\delta\psi\right].
 \eea
The Klein-Gordon equation for the two fields read
\be \label{kg_phi}
\delta\ddot{\phi}+2H\delta\dot{\phi} - \nabla^2\delta\phi = \dot{\phi}\left(3\dot{\Phi} + \dot{\Psi}\right) + \left(3H + 2\ddot{\phi}+H\dot{\phi}\right)\Psi
 - a^2\delta R,
\ee
\be \label{kg_psi}
\delta\ddot{\psi}+2H\delta\dot{\psi} - \nabla^2\delta\psi = 
\dot{\psi}\left(3\dot{\Phi} + \dot{\Psi}\right) + \left(3H + 2\ddot{\psi}+H\dot{\psi} + 2f''\phi^2\right)\Psi
+ 2f''\dot{\phi}\delta\dot{\phi}
+ \left(f'''\dot{\phi}^2-2f''a^2R\right)\delta\phi 
- 2f'a^2\delta R,
\ee
where the curvature scalar is
\be \label{ricci}
a^2 R =  6(\dot{H}+H^2) + 2\left[-3\ddot{\Phi} - 9\dot{\Phi} + 2\nabla^2\Phi - 3\dot{\Psi} - (6\dot{H}+6H^2+\nabla^2)\Psi\right].
\ee
The matter components then obey
\be \label{d_evol}
\dot{\delta}_m = 3H(w-c_s^2)\delta_m + (1+w)(-kv_m + 3\dot{\Phi}),
\ee
\be \label{v_evol}
\dot{v}_m = (3c_a^2-1)Hv_m+k\Psi+\frac{kc_s^2}{1+w}\delta_m - \frac{2k}{3(1+w)}\pi_m.
\ee
We consider later the forms of this system both in the vacuum and in the presence of matter.

\subsection{Vector and tensor perturbations}

The find out how the rotational perturbation evolves, we need only the field equation
\be
k^2{\bf W} = \frac{2a^2}{\kappa^2}(\rho+p)\frac{1}{1+\psi}{\bf v}_m,
\ee
and the conservation equation
\be
\left[a^4(\rho_m+p_m){\bf v}_m\right]^\bullet = -\frac{a^2}{\kappa^2}\boldsymbol{\pi}_m.
\ee
Combining these gives
\be
\dot{{\bf W}} + (2H+\frac{\dot{\psi}}{1+\psi}){\bf W} = \frac{1}{(1+\psi)\kappa^2}\boldsymbol{\pi}_m.
\ee
Thus the rotational perturbation can grow only in the case $d\log{(1+\psi)}/d\log{a}  < -2$. We can safely neglect 
the vector perturbations in these models, as usual in cosmology.

The evolution of the gravitational waves is given by
\be
\ddot{H}_{+,X} + \left(2H+\frac{\dot{\psi}}{1+\psi}\right)\dot{H}_{+,X} + k^2 H_{+,X} = \frac{2}{\kappa^2(1+\psi)}\pi^m_{+,X},
\ee
so the different polarizations evolve in completely same way. Thus the waves still propagate with the speed of light and 
are sourced only by anistropic stresses of fluids. The decay of the amplitude of a cosmological background of gravity waves 
is given by the expansion rate in a way which is only slightly modified by the evolution in $\psi$.

The propagation of scalar modes in these models turns out to be more involved, and will be considered in the separate section (\ref{sektio4}). 
 
\section{Newtonian limit}
\label{sektio3}

\subsection{Solar system}

Consider first the Newtonian limit of the cosmological perturbations to study the effects within our Solar system. 
In the Solar system we may neglect the cosmological expansion and set $a(\tau)=1$. Furthermore, one may assume 
the time derivatives are negligibly small with respect to space derivatives. We consider vacuum outside a mass and
thus can neglect the source terms. The ADM energy constraint (\ref{admenergy}) then reduces to
\be
\nabla^2\Phi = \frac{1}{2(1+\psi)}\nabla^2\delta\psi.
\ee 
Assuming spherical symmetry, the solution to this equation is $\Phi = \delta\psi + C_1/r + C_2$. Demanding that we recover
the usual Schwarzschild solution in the limit of GR fixes the two integration constants as $C_1 = -2GM$, $C_2=0$, where $M$ is
the mass of the spherical object. This shows that the gradient of the field $\psi$ acts as an extra source in the post-Newtonian 
generalization of the Poisson equation. The shear constraint (\ref{propagation}) shows that the gravitational potentials are not 
equal due to the effect of this field,
\be \label{as}
\Phi-\Psi = \frac{\delta\psi}{1+\psi}.
\ee
The Raychaudhuri equation (\ref{ray}) is redundant with the above constraints at this limit. We then exploit the Klein-Gordon 
equation (\ref{kg_psi}) to obtain the constraint 
\be \label{kg}
\delta\psi = -4f'(\phi)\left(\Psi-2\Phi\right),
\ee
where we have integrated again twice and killed the integration constants by requiring consistency at the GR limit. 
Now we can eliminate the gradient $\delta\psi$ from the above equations. The solution for the gravitational potentials is then
\be
\Psi =  
-\frac{2GM}{1+\psi}
\left(\frac{1+\psi-8f'}{1+\psi-6f'}\right), \quad
\Phi = 
-\frac{2GM}{1+\psi}
\left(\frac{1+\psi-4f'}{1+\psi-6f'}\right),
\ee
where the background values $\psi$ and $f'(\phi)$ are constants. 
So, we may write the metric as
\be
ds^2 = 
-\left(1-\frac{2G_*M}{r}\right)d\tau^2 + \left(1+\frac{2\gamma G_*M}{r}\right)dr^2 + r^2d\Omega^2, 
\ee
where the effective gravitational constant is
\be \label{g_eff}
G_* = \left(\frac{1+\psi-8f'}{1+\psi-6f'}\right)\frac{G}{1+\psi},
\ee
and the post-Newtonian parameter $\gamma$ is given as
\be \label{gamma}
\gamma = \left(\frac{1+\psi-4f'}{1+\psi-8f'}\right).
\ee  
There are very tight constraints on the time variation of $G_*$ and the deviation of $\gamma$ from its value in general relativity $\gamma=1$ \cite{Uzan:2002vq}.
As the current accuracy in determination reaches few parts in hundred thousand, the most stringent constrain ensuing the tracking of the Cassini 
spacecraft \cite{Will:2005va}, we can estimate on that our fields should satisfy $|f'(\phi)|, |\psi| \lesssim 10^{-4}$. 
If we would naively set the field $\phi$ to its cosmological value in the models where the 
$f(\phi)$ correction drives the present acceleration, those models would be immediately ruled out! However, since the exact Schwarzschild solution 
is Ricci-flat, $R=0$, for a properly regularized operator one should have $\phi \equiv \Box^{-1}R = 0$ there. Constraints can still be imposed. For this
purpose, expand the coupling $f(\phi)$ as a power series, 
\be
f(\phi) = \sum_{k=0}^\infty f_k\phi^k. 
\ee
Negative powers are excluded since we want Ricci-flat spaces to exist. We find that
\be
G_* = \left(\frac{1+f_0-8f_1}{1+f_0-6f_1}\right)\frac{G}{1+f_0}, \quad \gamma = \left(\frac{1+f_0-4f_1}{1+f_0-8f_1}\right).
\ee
Therefore it is possible to constrain the constant and the linear parts of the coupling from Solar system experiments. However, the constant $f_0$ may always
be absorbed into a redefinition of the $\kappa \rightarrow \kappa/(1+f_0)$ in the action (\ref{action}), and so one may set $f_0=0$. 
We then get the strict bound
\be \label{gammab}
-5.8 \cdot 10^{-6} < f_1 < 5.7 \cdot 10^{-6} 
\ee
Deviations from GR ensuing from the higher order terms in the expansion for $f(\phi)$ are of the second order in perturbation 
theory, $\sim (GM/r)^2$. Since the gravitational potential $\Phi = -GM_\odot/r_\odot \sim 10^{-6}$ at the surface of the Sun, we 
expect the order of magnitude of the nonlocal corrections of the type $f_2\phi^2+f_3\phi^3+\ldots$ in these models to be within the 
experimental bounds based on the Schwarzschild geometry. 

\subsection{Structure formation}

Next we consider cosmological scales and relax some of the simplifying assumptions. We then allow time evolution and do not 
impose spherical symmetry. We take into account pressureless matter source and thus focus on the formation of inhomogeneous 
structure in dust-dominated cosmology. We look at the first the subhorizon limit. Then the gradient terms are more important 
than the time derivatives of the inhomogeneities, $\ddot{y} \sim H^2 y \ll k^2 y$. This approximation could be called {\it the 
linear post-Newtonian limit}, as in \cite{Amendola:2005cr}.

Differentiating Eq.(\ref{d_evol}) and using Eq.(\ref{v_evol}) quickly gives us
\be \label{d_evol2} 
\ddot{\delta}_m + H\dot{\delta}_m = k^2\Psi.
\ee
We need to solve the right hand side of this equation.
The response of the gravitational potential to the matter source is now modulated by the gravity coupling,
\be \label{po1}
k^2\Phi = \frac{1}{2(1+\psi)}\left(\frac{a^2}{\kappa^2}\delta\rho_m + k^2\delta\psi\right).
\ee
The Raychaudhuri Eq. (\ref{ray}) at this limit gives 
\be \label{po2}
-k^2\Psi = \frac{1}{2(1+\psi)}\left(\frac{a^2}{\kappa^2}\delta\rho_m + k^2\delta\psi\right).
\ee
It is through these two relations that the modified gravity effects could now show up in the integrated Sachs-Wolfe effect and in the weak 
lensing potential. These effects and their correlation with the cosmic microwave background anisotropies could thus be used to probe these gravity models. The latter is 
relevant at the large scales, where the use observations is limited by the cosmic variance. From here we
see that the relative change of the potentials is of roughly of the order $\psi$, which is typically negative and order of $\mathcal{O}(0.1)$ in the simplest 
models accelerating the universe nowadays \cite{Koivisto:2008xf}. This is in a viable but interesting range. 
  
Combining Eqs. (\ref{po1},\ref{po2}) with Eq.(\ref{kg_psi}) gives us the gradient of $\psi$ in terms of $\delta\rho_m$. Then Eq.(\ref{d_evol2}) becomes
\be \label{d_evol3}
\ddot{\delta} + H\dot{\delta} = 4\pi G_* a^2\rho_m\delta = 
\left(\frac{1+\psi-8f'}{1+\psi-6f'}\right)\frac{4\pi G a^2}{1+\psi}\rho_m\delta_m,
\ee
where the expression for the effective gravitational constant is the same as in Eq.(\ref{g_eff}). Now it is, unlike in the near-Schwarzschild 
approximation, in general time-dependent, if the background $\phi$ and $\psi$ are evolving. A crucial feature here is the absence of
effective sound speed terms. These are known to appear in some coupled dark energy \cite{Koivisto:2005nr} and in modified gravity \cite{Koivisto:2005yc}, causing
then instability in the dynamics of linear perturbations. This can be tightly constrained by observations of the galaxy distribution. However, our result
(\ref{d_evol3}) suggests that the growth of structure is here scale-independent, and thus the constraints will now ensue solely from the overall 
normalization
and the growth rate. Since the $G_*/G$ is of order one or less in the $f(\phi)$ dark energy models, we expect of at most of order one modifications to the
growth rate at late times. This is within the observed limits at the present \cite{Verde:2001sf}, but could be a useful additional constraint on these 
models.   

Finally we remark that all of the results about the large-scale structure formation in this subsection rely crucially on the assumption that dark matter 
is exactly cold (a perfect and pressuless fluid). By relaxing this assumption one may drastically change the predictions of modified gravity models concerning 
the inhomogeneities, while keeping both the gravity sector and the background expansion fixed \cite{Koivisto:2007sq}. 
        
\section{Ghosts, causality and stability}
\label{sektio4}

To study the propagation of the scalar modes, it is useful to go to the Einstein frame.
Perform the Weyl rescaling \cite{Wands:1993uu,Faraoni:2006fx}
\be \label{weyl}
\tilde{g}_{\mu\nu} \equiv (1+\psi)^{\frac{2}{D-2}}g_{\mu\nu} \equiv e^{\frac{2}{D-2}\lambda}g_{\mu\nu}.
\ee
For a moment we consider the $D$-dimensional case.
The action then becomes \cite{Koivisto:2008xf}
\be
\label{se}
S  =  \int d^D x \sqrt{-\tilde{g}}\left[\tilde{R} - \frac{n-1}{n-2}(\tilde{\nabla}\lambda)^2 -
e^{-\lambda} f'(\phi)(\tilde{\nabla}\phi)^2 + \tilde{\nabla}_\alpha\lambda\tilde{\nabla}^\alpha\phi  +
2\kappa^2e^{-2\lambda} \mathcal{L}_m(\tilde{g}e^{-\lambda}). \right]
\ee
Now $\psi$ has also acquired a kinetic term. The curvature coupling has been removed in this frame, but the matter sector has now become 
nonminimal. However, this does not complicate our analysis if solely the vacuum is considered. Furthermore,
it is possible to to diagonalize the kinetic Lagrangian to get rid of the derivative interaction. For that
purpose, we must define the new field $\xi$
\be
\xi \equiv \sqrt{\frac{D-1}{D-2}}\left(\lambda - \frac{D-2}{2(D-1)}\phi\right).
\ee
The diagonalized action is then in vacuum 
\be
\label{se2}
S  =  \int d^D x \sqrt{-\tilde{g}}\left[\tilde{R} - \frac{1}{2}(\tilde{\nabla}\xi)^2 -
\frac{1}{2}\left(e^{-\sqrt{\frac{D-2}{D-1}}\xi-\frac{D-2}{2(D-1)}\phi}f'(\phi)
- \frac{D-2}{4(D-1)}\right)(\tilde{\nabla}\phi)^2 
%+2\kappa^2e^{-2\sqrt{\frac{n-2}{n-1}}\xi-\frac{n-2}{n-1}\phi)} \mathcal{L}_m(\tilde{g}e^{-\sqrt{\frac{n-2}{n-1}}\xi-\frac{n-2}{2(n-1)}\phi})
. \right]
\ee
In presence of (non-conformal) matter both fields would now be non-minimally coupled to the matter sector. 
One notes already that in the $f \rightarrow 0$ limit we have one canonic and one phantom scalar field (except in $D=2$ where the phantom field 
disappears). Indeed we will find that this is the usual field content of these models also in the more general case.  
In the rest of this section,
we omit the tildes from the Einstein frame quantities to ease notation, since every variable will be evaluated in the Einstein frame except when explicitly mentioned.

When analyzing perturbations in multi-field systems it is convenient to perform a decomposition of the perturbations into adiabatic 
and entropic parts \cite{Gordon:2000hv}. In our two-field case the entropy will then be presented by just one field. Except for 
possible differences in signs, our form (\ref{se2}) belongs to the non-linear sigma models studied in \cite{Koshelev:2005wk}. 
For a recent consideration on more general non-linear sigma models, see \cite{Langlois:2008mn}. To proceed, set now $D=4$ and call 
\be
F(\phi,\xi) \equiv e^{-\sqrt{\frac{2}{3}}\xi-\frac{1}{3}\phi}f'(\phi) - \frac{1}{6}. 
\ee
The background adiabatic field $\sigma$ is implicitly given by 
\be
\dot{ \sigma } = \sqrt{|F \dot{\phi}^2 + \dot{\xi}^2|},
\ee
and the entropy field is orthogonal to this in the field space. 
The entropic and adiabatic perturbations can then be written as
\bea \label{def3}
\delta      s & = & \frac{\sqrt{|F|}}{\dot{\sigma}}\left(\dot{\phi}\delta\xi - \dot{\xi}\delta\phi\right), \\ 
\delta \sigma & = & \frac{1}{\dot{\sigma}}\left(\dot{\xi}\delta\xi + \sqrt{|F|}\dot{\phi}\delta\phi\right).
\eea
With the four previous definitions, the Klein-Gordon equations for the fields $\phi$ and $\xi$ and a lot a algebra, one obtains the simple results 
\be \label{sigma_evol}
\ddot{\delta\sigma} + 2H\dot{\delta{\sigma}} + k^2\delta\sigma = \dot{\sigma}\left(\dot{\alpha} - 3H\alpha + a\kappa\right),
\ee
\be \label{s_evol}
\ddot{\delta s} + 2H\dot{\delta s} + \left[\left(-\frac{1}{2}\frac{F_{,\xi\xi}}{F} + \frac{1}{4}\frac{F_{,\xi}^2}{F^2}\right)\dot{\sigma}^2 + 
k^2\right]\delta s 
= 0.
\ee
Thus we have found the two independently evolving scalar degrees of freedom of these models. 
Coupling of the evolution equations seems to be a almost generic property of generalized multi-field models \cite{Langlois:2008mn}.
The fact that the adiabatic and entropy field evolutions now decouple is due to the masslessness of the fields (they do not have potentials, 
self or otherwise), and the diagonalizability of the kinetic Lagrangian in the Einstein frame. 
One may choose a gauge where the left hand side of Eq.(\ref{sigma_evol}) vanishes, but in general the perturbations of the $\sigma$-field
are sourced by the metric perturbations. The evolution equation for the entropy mode, on the contrary, is homogeneous. This means that 
if the $s$-field is initially smooth, it will stay so. At small scales any initial perturbation will be washed away because of the sound
speed equals unity, but at large scale there if a potential instability if $\sqrt{F}_{,\xi\xi}/\sqrt{F}>0$. Most importantly, also the 
other field has the propagation speed $c_A^2$ equal to the light speed, just like it is for the canonical scalar fields, $c_A^2=1$. 
Therefore, in all directions of the field space the propagation speed is unity. Since the Weyl rescaling retains the light cones,
the scalar perturbations propagate in both frames with the speed of light. Thus we can a posteriori justify performing the
analysis in the Einstein in this case, though in general the propagation velocity of perturbations may not be frame-invariant. 

Thus, at the classical level there is no instability (since $c_A^2>0$) or causality violation (since $c_A^2 \le 1$). 
At the quantum level, an inconsistency may occur if the kinetic terms of the fields have a wrong sign. 
Our derivation above assumed that $F(\phi,\xi) \ge 0$. If this is not case, the kinetic term of $\phi$ would have the wrong sign, 
implying an appearance of a ghost at the quantum level. If furthermore $F(\phi,\xi) < (\dot{\xi}/\dot{\phi})^2 $, the adiabatic field direction
in the field space corresponds to a ghost degree of freedom. Translated into a condition for the original Jordan frame fields,
the no-ghost requirement for the scalar modes reads simply
\be \label{noghost}
\frac{f'(\phi)}{1+\psi} - \frac{1}{6} > 0 \Rightarrow 6 f'(\phi) > 1+\psi,
\ee 
where in the second line we have used the tensor no-ghost requirement.     
% One notes that the consistency of the model tends to require the acceleration. 
% In the exponential and power-law models late-time effects to occur iff the correction function grows with the 
% field, $f'(\phi)>0$, which is a necessary condition for the absence of ghosts. The time derivative of the correction function $f$ is however 
% negative and in models which feature to acceleration the effective gravitational coupling $1+\Psi$ must decrease, and ensures that the
% inequality (\ref{noghost}) will satisfied also at late times. 
In fact, the effective equation of state of the (Jordan frame) universe can be expressed as
\be \label{w_eff}
w_{eff} = \frac{3w_m\Omega_m + f'(\phi)(\frac{1}{2}\frac{1}{H^2}\dot{\phi}^2-6) + f''(\phi)\frac{1}{H^2}\dot{\phi}^2
- (1-\frac{1}{2H}\dot{\phi})\frac{1}{H}\dot{\psi}}{3(1+\psi-6f'(\phi))}.
\ee
Note that this is fully general expression and does not assume vacuum.
Thus the future sudden singularity occurs just when the limit (\ref{noghost}) would be crossed. Is then clear that kinetic term of the scalar mode
does not change its sign. However, we can make the observation that exactly when the inequality (\ref{noghost}) becomes an equality, i.e. the scalar 
kinetic term changes it's sign, the gravitational constant (\ref{g_eff}) diverges. In fact this applies also for the tensor mode kinetic term, since $G_* 
\sim 1/(1+\psi)$. The same phenomena has been observed to occur in Gauss-Bonnet cosmology. There exist peculiar instabilities, of 
stringy vacua \cite{Charmousis:2008ce} and of linear tensor perturbations \cite{Kawai:1998ab, Kawai:1999pw}. In particular, a 
divergence of scalar modes has been associated with a crossing of a ghost limit \cite{Koivisto:2006xf,Koivisto:2006ai}. There, in contrast to our 
present case, the background can continue to evolve smoothly while the singularity occurs only at the linear level. 

\section{Conclusions}
\label{sektio5}

We considered a class of cosmological models based on the generalized gravity action (\ref{action}). Though apparently nonlocal, this action may be recasted 
into a local biscalar-tensor theory. The theories represent a subclass of more general covariant theories, involving more, possibly infinite number of 
scalar fields. The simple parameterization of the corrections to Einstein gravity as the function $f(R/\Box)$ has already several new theoretical features and 
potential to describe new phenomenology. In particular, one may construct dark energy cosmologies without introducing unnatural parameters and explain
the coincidence of the similar matter and dark energy densities just today by the response of $R/\Box$ to the onset of matter domination.

In this paper we focused on the cosmological perturbations and weak field limit of these models. The scalar, vector and tensor perturbation equations
were derived up to linear order in the flat FRW background. To simplify the equations, we fixed a gauge for each type of perturbation, but of course
the equations in an arbitrary gauge may be obtained from our results using the standard transformation formulas. The scalar equations we wrote down
in the Newtonian (or longitudinal) gauge. We considered the evolution of inhomogeneities in a matter dominated universe. We found that gradient effects are
absent, but that the growth rate is modified according to a time evolving effective gravitational constant. Also, as rather generic predictions of 
modified gravity, the Poisson equation is modified and there is effective anisotropic stress. These effects might provide hints to detect these models from
any observed deviation from the general relativistic predictions. In particular, the integrated Sachs-Wolfe effect probes the late time variation of the 
gravitional potentials \cite{Jain:2007yk,Bertschinger:2008zb}, and the small scale effects of modified gravity might be probed by weak lensing 
experiments \cite{Amendola:2007rr,Schmidt:2008hc}.
We hope to quantify these constraints on specific models by a detailed numerical analysis in a future work. In addition, inflation and 
the generation of the primordial spectrum would be interesting to study with the action (\ref{action}). 

We considered here the Post-Newtonian limit for static spherically symmetric solutions, which is crucial for the local tests of general relativity. We found 
that there are two PPN type parameters possibly differing from their general relativistic values, with a given dependence of the form of $f$. This 
dependence allows to constrain tightly the constant and the linear parts of the function $f$. In particular, they are severely restricted by the bounds on the 
post-Newtonian $\gamma$ obtained from the tracking of the Cassini spacecraft, resulting in Eq.(\ref{gammab}). For example, the purely linear model 
resulting in $\Omega_m = 0.3$ today 
is ruled out by this constraint alone (though this 
model would not accelerate the universe enough anyway \cite{Koivisto:2008xf}), as well as the simplest linear stabilator of the Euclidean action 
(though it already had tension with the Big Bang nucleosynthesis \cite{Wetterich:1997bz}). The constraints on the higher powers in $f$ seem to be 
some orders of magnitude looser.

The issues of causality and stability were also considered. The rotational perturbations were shown to decay as usually. The gravitational waves evolve
with an extra friction term and a modulated response to the matter stress sources due to the coupling to $(1+\psi)$. The evolution of the two extra scalar 
degrees of freedom is more delicate. We found that it is possible to separate these degrees of freedom into decoupled modes by considering the theory in the 
Einstein frame, diagonalizing the kinetic terms and then considering the perturbations $\delta s$ and $\delta \sigma$, which are fluctuations of implicitly 
defined combinations of the fields appearing in the original Jordan frame Lagrangian. However, since it is possible to write down decoupled, canonical actions 
for those two perturbations, we believe it is legitimate to regard them as the physically propagating degrees of freedom. This recipe would not probably work 
in a more general case though. Here our result is that the new scalar modes propagate perturbations with the light speed, and thus they are causal and stable. 
The formal ghost conditions are directly linked to divergence of classical perturbations 
(similar to the Gauss-Bonnet cosmology) and here also with the sudden future singularity of the 
cosmological background. Thus these conditions clearly indicate a pathological point of the solutions. We note that the singularity 
structure of the $f(R)$ models has also raised recent concerns \cite{Frolov:2008uf,Kobayashi:2008tq,Dev:2008rx,Bamba:2008ut}. 

To conclude, the $f(R/\Box)$ gravities seem have a viable Newtonian limit, although they may feature significant modifications of the cosmological 
expansion.

\acknowledgments
I thank Chris Byrnes and Subodh Patil for useful discussions.

\bibliography{nl2}

\end{document}